# Tailored Fabrication of 3D Nanopores with Dielectric Oxides for Multiple Nanoscale Applications

*German Lanzavecchia[1,2], Anastasiia Sapunova[1,3], Ali Douaki[1], Shukun Weng[1,3], Dmitry Momotenko[4], Gonçalo Paulo[5], Alberto Giacomello[5], Roman Krahne[1*], Denis Garoli[1,6*]*

[1] Istituto Italiano di Tecnologia, Via Morego 30, Genova, Italy

[2] Dipartimento di Fisica, Università degli Studi di Genova, Via Dodecaneso 33, 16146, Genova, Italy

[3] Università degli Studi di Milano-Bicocca, Piazza dell'Ateneo Nuovo, 1 - 20126, Milano, Italy

[4] Institute of Chemistry, Carl von Ossietzky Universität Oldenburg, Oldenburg D-26129, Germany

[5] Dipartimento di Ingegneria Meccanica e Aerospaziale, Sapienza Università di Roma, Via Eudossiana 18, 00184 Roma, Italy

[6] Dipartimento di Scienze e Metodi dell'Ingegneria, Università degli Studi di Modena e Reggio Emilia, Via Amendola 2, 43122, Reggio Emilia, Italy

Email: denis.garoli@unimore.it; roman.krahne@iit.it

**ABSTRACT**

Nanopore sensing is a key technology for single-molecule detection and analysis. Solid-state nanopores have emerged as a versatile platform, since their fabrication allows to engineer their properties by controlling size, shape, and chemical functionalization. However, lithography-based fabrication approaches for non-planar nanopores-on-chip rely on polymers that have limits with respect to hard- and robustness, durability, and refractive index. In this respect, nanopores made of metal oxides with high dielectric constant would be much more favourable and have the potential to extend the suitability of solid-state nanopores towards optoelectronic technologies. Here, we present a versatile method to fabricate three-dimensional nanopores of different dielectric oxides with controlled shapes. Our approach uses photoresist only as a template in the focused-ion-beam lithography to define the nanopore shape, which is subsequently coated with different oxides ($SiO_2$, $Al_2O_3$, $TiO_2$ and $HfO_2$) by atomic-layer deposition. Then the photoresist is fully removed by chemo-physical treatment, resulting in nanopores entirely made from dielectric oxides on a thin solid-state membrane. Our methodology allows straightforward fabrication of convex, straight, and concave nanopore shapes that can be employed in various technologies and applications. We explored their performance as ionic nanochannels and investigated the dependence of the ionic current rectification on the nanopore geometry. We found hysteresis in the ionic conductance that enables potential applications of the nanopores in memristors. We also investigated the dielectric oxide nanopores for DNA sensing by measuring both cis-trans and trans-cis translocations and support our data with numerical simulations based on the Poisson-Nernst-Planck model. These results, showcasing different shapes and oxides, confirm the robustness and tunability of the nanopores and their suitability for different technologies. We note that the presented fabrication route can be extended to a large range of 3D shapes to realise versatile nanostructures using a variety of dielectrics.

**Introduction**

Over recent decades, key advancements in nanopore technology have paved the way for improved applications in various fields, including sequencing[1], biosensing[2,3], nanofluidics[4], ion transport [5] and selectivity[6], electroosmosis[7], energy harvesting[8] and electronic devices[9,10]. Notably, nanopores offer direct physical access to biomolecule analysis[11,12] and play a pivotal role in third-generation sequencing[13] and DNA data storage[14]. However, challenges persist in nanopore technology, including limited pore lifetime and stability[15], difficulties in tuning pore size[16], limited selectivity, and scalability concerns from single pores to pore arrays configurations. Solid-state nanopores[17,18], in particular, have become the focus in single-molecule sensing due to their high robustness and durability[19], even under harsh conditions in terms of temperature, pressure, and pH. Various fabrication alternatives have been developed[20], utilizing a wide range of materials, wherein the shape and size of the nanopores are defined by the fabrication method[8].

The geometric distribution of charges inside the nanopore, determined by the surface charge of the material and by the 3D shape of the nanochannel[21,22], can have important effects on the nanopore's functionality. Depending on the shape and material of the nanopore, the inner surface charges influence the ion concentration and electrostatic potential inside the nanopore[23.] This leads to Ionic Current Rectification (ICR)[24,25], with highly asymmetric nanopores producing high ICR ratios[26]. Moreover, the nanopore conductance can depend on the behavior of the surface charges that redistribute within the nanopore channel due to the finite mobility of ions under applied potentials[27,28] which leads to memristive behavior, as discussed in some recent works[29,30].

The choice of material and geometry can also influence the noise in nanopore ionic current measurements[17,31,32]. For example, quartz nanopipettes have been employed to measure the current blockade of geometrical features separated only 6 nm along a DNA strand[33], attributing this super-resolution to the enhancement of the electric field at the tip of the nanopore. However, a quartz nanopipette limits the application to a single nanopore. On the contrary, nanopores on a silicon chip can [34] be fabricated as nanopore arrays[35] for parallel and high-throughput applications. In this view, $HfO_2$ step-like conical nanopore arrays on $Si_3N_4$ membranes have been recently fabricated for biosensing and energy harvesting, using a modified Atomic Layer Deposition (ALD) configuration [8]. Taking inspiration from this work, we present a robust method to fabricate three-dimensional on-chip nanopores with arbitrary geometries, made of different dielectric oxides. These structures can be readily realized on solid-state substrates (on-chip) and fabricated as single pores, as pore arrays or in arbitrary arrangements. The fabrication is based on focused ion beam (FIB) lithography that produces a hollow conical structure of a photoresist mold onto a silicon/silicon nitride chip[36.] Then the metal oxide is deposited from the back side of the membrane by ALD or physical vapor deposition (PVD)[37]. The crucial step is the full removal of the hardened photoresist that we achieved by UV light exposure (to break the crosslinking of the polymer chain in the resist) followed by low-power oxygen plasma in our Reactive Ion Etching with Inductively Coupled Plasma (ICP-RIE) setup, or by heating the samples to 600 °C under O2 atmosphere. With this process we obtain nanopores with few nm apertures consisting only of the metal oxide, while their shape can be controlled in 3D by the FIB lithography[38]. We demonstrate the fabrication of nanopores from four different oxides ($SiO_2$, $Al_2O_3$, $TiO_2$, and $HfO_2$), and three different shapes (with concave, convex, or straight edge profiles) and investigate their performance in ICR, single molecule detection and as ionic memristors.[3927]

**Results and discussion**

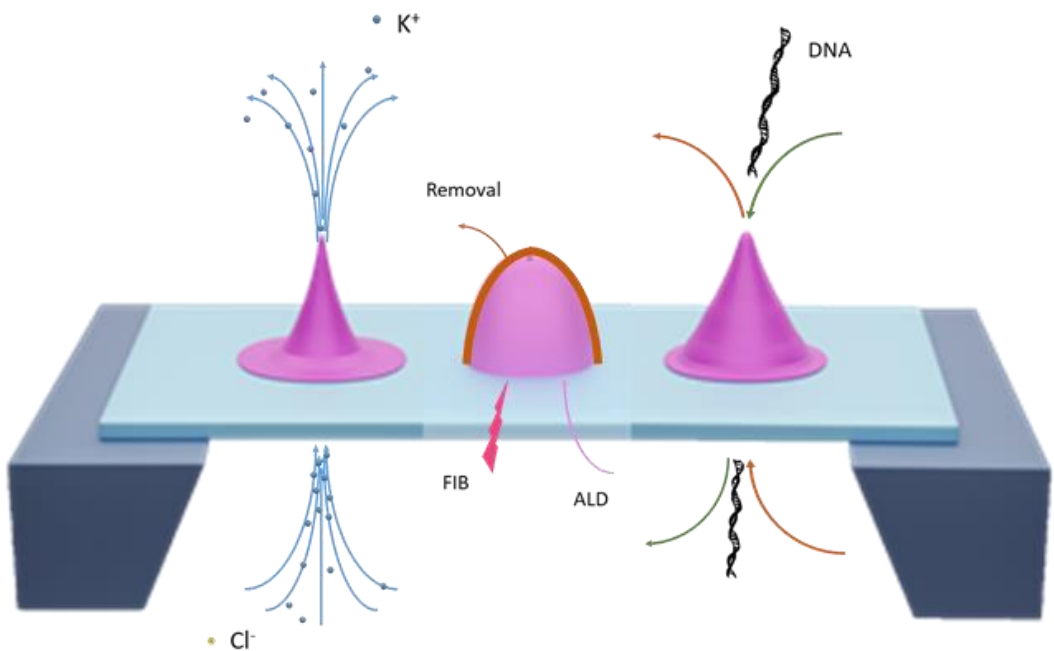

**Figure 1**. Schematic of the 3D nanopore concept.

*Nanopore fabrication*

The fabrication process is outlined in Figure 1 and some examples of the fabricated 3D nanopores are illustrated in Figure 2 (in the Supporting Information we reported several additional examples of single nanopore and nanopores array to demonstrate the versatility of the method). The key features are the deposition of the photoresist from the top side, with the successive FIB lithography and metal oxide deposition from the backside of the membrane. This combination retains the freedom in the design of the nanopore of the FIB, and enables the full removal of the photoresist mold after metal oxide deposition. An additional advantage of the approach is that the nanopore aperture can be defined with nanometer resolution by the fine control of the ALD of the dielectric metal oxide. To fabricate 3D hollow conical structures with different edge profiles in the photoresist, we adapted a technique based on FIB milling of concentric disks with different diameters, as illustrated in Figure 1 and Figure S8 [36,40,41]. Here the secondary radiation and electron scattering related to the ion-beam exposure results in crosslinking of the photoresist in the vicinity of the removed disk volume, which results a layer of hardened photoresist with ca 50 nm thickness that wraps the external shape of the disk stacks. This photoresist mold is then used as a template for the dielectric oxide nanopores.

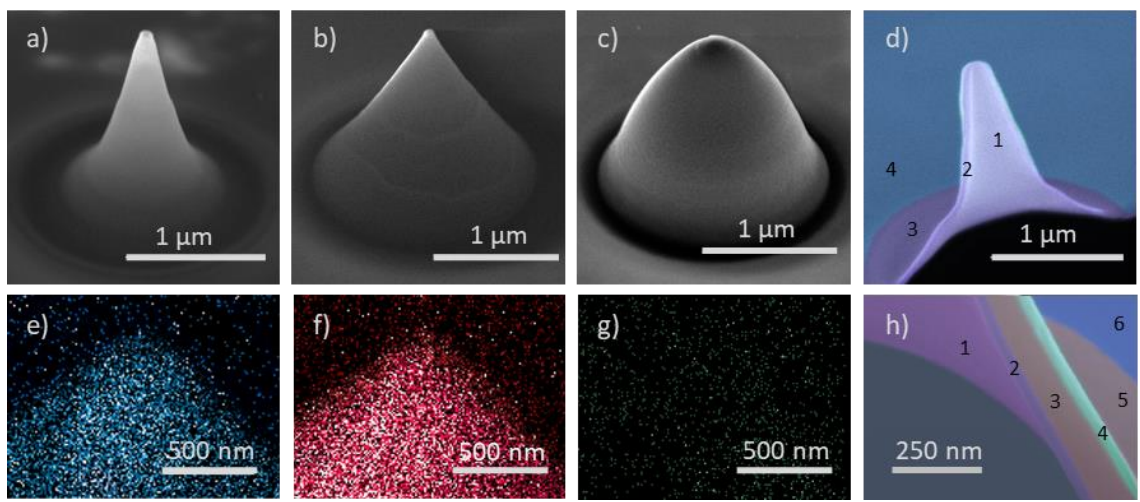

**Figure 2**. Electron microscopy analysis of metal oxide nanopores with different shapes. SEM micrographs of $Al_2O_3$ nanopores with concave (a), straight (b), and convex (c), edge profiles. The aperture of the nanopores is around 30 nm in all cases. (d) Cross section of a concave $Al_2O_3$ nanopore (marked 1- interior wall of the nanopore in $SiO_2$, 2- cross section of the dielectric wall, 3- outer $SiO_2$ wall, 4- $Si_3N_4$ membrane). (e-g) Compositional analysis of an Al2O3 nanopore by electron dispersive spectroscopy (EDS) mapping, showing the signals for Al (e), O (f), and C (g). (h) Cross section of a convex $SiO_2$ nanopore before photoresist removal, the $SiO_2$ and photoresists section are highlighted by false colors (marked 1- interior wall of the nanopore in $SiO_2$, 2 – cross

section of the dielectric wall, 3- cross section of the photoresist wall, 4- Pt layer, 5- outer photoresist wall, 6- Si$_3$N$_4$ membrane).

We note that we distinguish here two different photoresist thicknesses: one is the overall layer thickness after spin coating that is in the micrometer range and which defines the height of the 3D nanopore, the other is the thickness of the crosslinked layer that defines the wall thickness of the nanopore in the mold. The wall thickness is reduced to ca 30 nm by an oxygen plasma cleaning that we apply before the metal oxide deposition. The metal oxide can be deposited by PVD or ALD. PVD can be directly applied to the samples since the process is directional, and therefore covers only the backside of the membrane and the photoresist cones. However, PVD films are slightly grainy, and therefore this approach provides only limited control on the nanopore diameter, with the risk of blocking the opening. With ALD very homogeneous layers with high control on thickness can be obtained[38], but it is important to assure the one-sided deposition in order to still be able to remove the photoresist mold. To achieve this, we place the sample top side down on a Polydimethylsiloxane (PDMS) substrate that adheres to the photoresist and protects the top side from the ALD, while leaving the back side accessible for metal oxide deposition. Thermal ALD is preferred over plasma ALD, as it produces higher quality oxide layers[42,43], and does not contribute to further hardening of the photoresist that would make the resist removal more difficult.

*Photoresist removal*

The removal of the hardened photoresist mold is challenging[44,45]. Standard oxygen plasma processes do not work as they only reduce the thickness[46] to a persistent layer of around 30 nm. Solvents like N-Methyl-2-pyrrolidone (NMP), designed to remove hardened photoresist, prove ineffective in the complete elimination of this layer. We tested piranha solution as an additional etchant, but this resulted in membrane damage, and aggressive oxygen plasma treatments performed in a RIE-ICP configuration led to damage of the thin dielectric layers. We found that resist removal using UV light to break the crosslinking in the photoresist[44,46], and then low-power oxygen plasma in an ICP-RIE leads to fast and effective removal of the photoresist. Another successful strategy that we applied for resist removal is thermal annealing at 650°C in the presence of oxygen, which facilitates the decomposition and degassing of organic materials[47], and concurrently improves the mechanical and optical properties of the oxide layer[48,49]. Both methods proved to be effective in fully removing the hardened photoresist layer, as confirmed by EDS mapping shown in Figure 2e-g.

*ALD – different materials*

The metal deposition by ALD on a mold gives access to a wide variety of dielectric oxides as materials for the 3D nanopores. We tested all materials available in our ALD system, which are $SiO_2$, $Al_2O_3$, $TiO_2$ (titanium dioxide), and $HfO_2$ (hafnium dioxide), see Figure 2 and Figures S4 & S5 in the SI. This set of dielectric oxides outlines the flexibility of our fabrication that allows to select the nanopore material to meet specific application requirements, in particular towards high dielectric constant materials that can open up possibilities to integrate nanophotonic resonators and Mie-tronics in the nanopore platform[50,51].

Now we turn to explore the properties and functionality of the all-dielectric oxide nanopores in a range of applications, that are ICR, single molecule detection, and as memristors. In this respect, we take advantage of the design freedom that our fabrication enables, and investigate the impact of the different oxide materials (SiO$_2$, Al$_2$O$_3$), different nanopore shapes (with concave, convex, and straight profiles), and pore diameters (ranging from 7-70 nm).

*Ionic current rectification*

ICR depends on the nanopore shape, size, and surface properties[52]. Figure 3a shows the conductance of Al$_2$O$_3$ nanopores that have different shape, but a similar channel length of 1.4 μm and pore diameter of around 70 nm in 10 mM KCl electrolyte. ICR is observed for all three shapes, with a higher current at positive bias with respect to negative bias, and this effect is strongest for the pores with straight profiles. We observe a similar behavior for SiO$_2$ nanopores with about 40 nm pore diameter, as depicted in Figure 3b. As could be expected 3D nanopores exhibit strong rectification with typical ICR ratios in range between 4.6 and 1.6, which is significantly higher than ICR in planar pores with ICR ratios around 1.

Our smallest nanopores with opening diameters of around 7 nm exhibit even stronger ICR. Figure 3c exemplifies the I-V curves for straight pore shapes, fabricated with SiO$_2$ and Al$_2$O$_3$. In such narrow nanopore the movement of the ions is dominated by the surface charge, thus the rectification effects are enhanced, and we obtain ICR at 500mV ratios of 7.0 for SiO$_2$ and 4.9 for Al$_2$O$_3$, whereas values reported for conical nanopores in a SiO$_2$ membrane are in the range of 1-2[53].

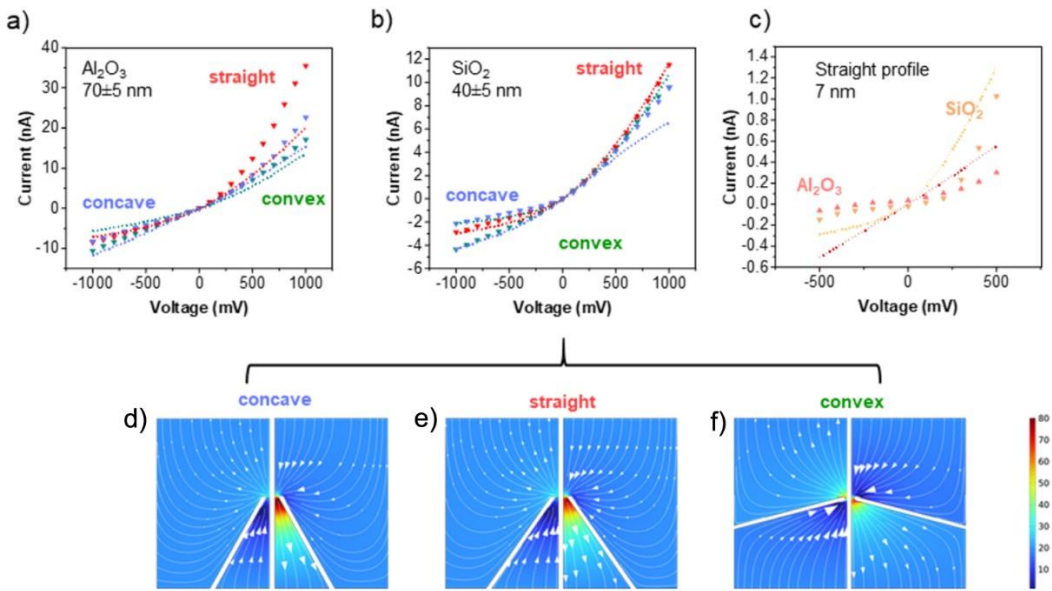

**Figure 3**. Ionic current rectification: measured (triangles) and simulated (dotted lines) I-V characteristics of the three nanopore geometries in 10 mM KCl electrolyte . a-c) current voltage curves for nanopores with different shapes and materials as described in the legends. d-f) Simulated concentration profiles of Cl⁻ [mol/m$^3$] and total flux (streamlines) for a SiO$_2$ nanopore with 40 nm diameter with d) concave, e) straight, and f) convex profile.

*Numerical simulations*

To elucidate the fact that straight pore profile features stronger ICR than the concave or convex shapes we performed finite element method modeling (COMSOL Multiphysics version 5.3). The Poisson-Nernst-Planck (PNP) equations[54,55] were employed to describe distributions of electric potential and ionic species concentrations across the entire domain as well as to simulate respective ionic fluxes . Further details can be found in the Supporting Information.  As shown in Figure 3d-f, which depicts the magnified view on the pore tip region for concave, straight and convex geometries, the major difference between the shapes is the value of the half-cone angle in the vicinity of the opening. This angle typically reaches 15-30 degrees for concave, 30-45 for straight and 80-70 for convex nanopores. The angle has a pronounced effect on local ionic distributions and ionic conductivity. As shown, as the angle of the nanopore increases from concave to straight, and then to convex shapes, one can observe a corresponding reduction in the conductivity asymmetry at opposite bias polarities. This consequence of a smaller degree of accumulation and depletion of ionic species at the tip of the nanopore at higher cone angles is related to a reduced asymmetry of the mass transport between the inner and outer parts of the nanopore. Inside the pore the mass-transport is rather restricted thus allowing stronger ion accumulation or depletion, depending on bias polarity, and is strongly dependent on the cone semi-angle whereas the mass transport outside the pore remains majorly unchanged. These variations in the geometry cause a major effect on ICR[21,56,57], with stronger mass-transport asymmetry (inside vs outside of the pore) that causes more intense high and low conductance states at opposite biases for concave, rather than straight, or even convex pore shapes.

Further differences between the fabricated nanopores are related to the magnitude of the surface charges, that are directly related to the choice of the pore material. The charge can be manipulated by nanopore fabrication methods, chemical functionalization of the pore walls as well as the interactions between the surface groups with ions in the electrolyte solution due to acid-base (pH) or other equilibria[21,53]. In this regard the main difference between $Al_2O_3$ and $SiO_2$ nanopores in relation to ICR lies in their surface charge. In our ICR simulations, we assigned surface charges from $–0.02$ to $–0.04$ $C/m^2$ and from $–0.06$ to $–0.1$ $C/m^2$ for $SiO_2$ and $Al_2O_3$ nanopores, respectively, thus resulting in a different degree of ICR. Therefore, the nanopore fabrication concept developed in this work offers a versatile platform to manipulate ICR in a wide range of conditions, from geometry to nanopore materials.

*DNA translocations*

Next, we discuss the properties of the nanopore in single molecule sensing, implemented by the electrical detection of the translocation of single λ-DNA molecules in 1M KCl solution through the nanopore in a nanopore reader[58].

The conical shape improved the overall signal-to-noise ratio compared to our previous work[59] with cylindrical nanopore geometry. The conical shape resulted also in an asymmetry with respect to the direction of translocation both in terms of frequency of events and signal to noise, that is the relative signal was $\Delta I/I_0 = 8\%$ for base-to-tip and $\Delta I/I_0 = 11\%$ for tip to base translocations (Figure 4a), and we observed approximately twice the amount of translocation events in the tip-to base direction with respect to the base-to-tip direction (Figure 4b). This is in agreement with recent results reported in HfO$_2$ step-like conical nanopores,[8] and glass nanopipettes[33].

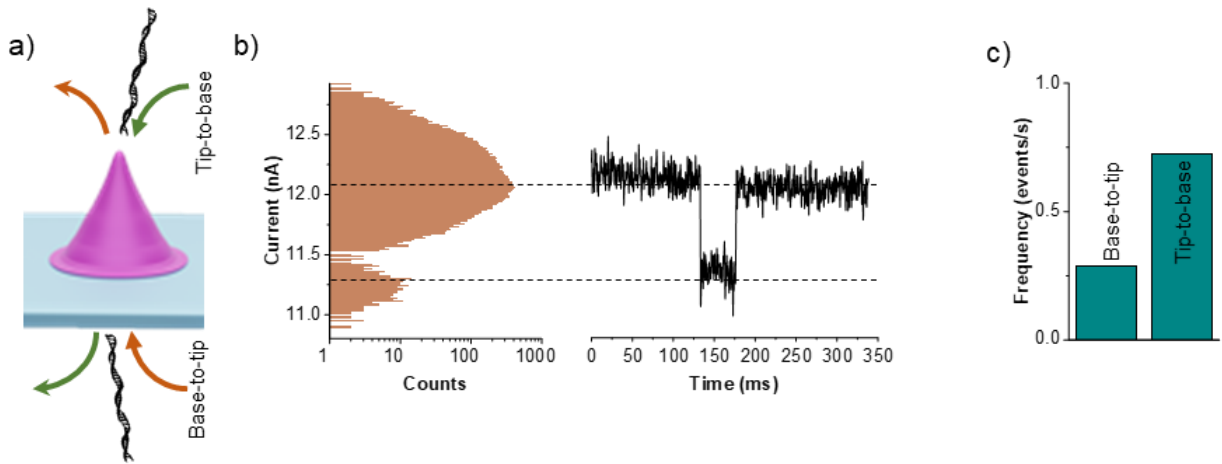

**Figure 4.** a) Schematic layout of the DNA translocation experiments b) Current blockade histogram of λ-DNA translocations through a straight profile nanopore in 1 M KCl at V = 500 mV,

measured at 20 kHz, with a characteristic translocation event on the right. c) Relative frequencies of DNA translocations in tip-to-base and base-to-tip directions.

*__Memristive behavior__*

Ionic memory resistors (memristors) are promising groundbreaking advancements in neuromorphic technologies[60–62] exploiting their capability to unite computing and memory in the same element. The conductance of known ionic memristors is voltage-dependent meaning that, at any given voltage, the conductance bears memory of the recent history of bias voltages across the ionic memristor. The memory effect observed in memristors is reminiscent of short-term synaptic plasticity in neurons, where the response of a neuron after a synapse depends on the history of the activations before the synapse[63,64].

Figure 5 reports the I-V curves of cyclic voltammetry in conical 3D nanopores. These plots show the pinched hysteretic behavior that is the fingerprint of memristors[65]. Figure 5a reports the hysteretic behavior of the 3 types of conical pores. Although there is significant variance between each shape, we observe that the conductance of the pores is higher if the (absolute) bias voltage was lower and that it decreases as higher (absolute) bias voltages are applied. For all cases, these pores behave as unipolar memristors, having a conductance-voltage curve that crosses close to 0, see figure 5b. This memristive behavior is similar to that reported in the literature for larger, concave pores[9,66]. The origin of the hysteretic behavior has its origin in the transient concentration polarization that is due to the dual asymmetry of the pores, their conical shapes and their charged surfaces, which takes place over diffusive timescales. Significant theoretical work has been done to estimate the expected memory timescale of the hysteretic loop, which should scale as $\propto \frac{L^2}{D}$, where L is the length of the pore and D the ionic diffusivity coefficient[67,68]. Using L=1.4 $\mu m$ and D= 1.4 $\frac{\mu m^2}{ms}$, wields timescales on the order of milliseconds, compatible with our results. As expected, we observe that the hysteresis curve depends on the frequency of the voltage cycle, observing larger loops for 12.5 mHz compared to 50 mHz, see figure 5c.

As an additional characterization of the memristor behavior, we show, in Fig. 4d, an experiment in which the nanopores are subjected to a number of positive voltage pulses followed by negative ones. The pulses produce a synaptic-like depression of the conductance with both positive and negative pulses consistent with a unipolar memristor. This means that the nanopore effectively "remembers" the previous pulses and that its downstream conductance can be "programmed" by providing an appropriate train of pulses. This behavior suggests that nanopores of this kind could be employed as elements of nanofluidic neuromorphic circuits.

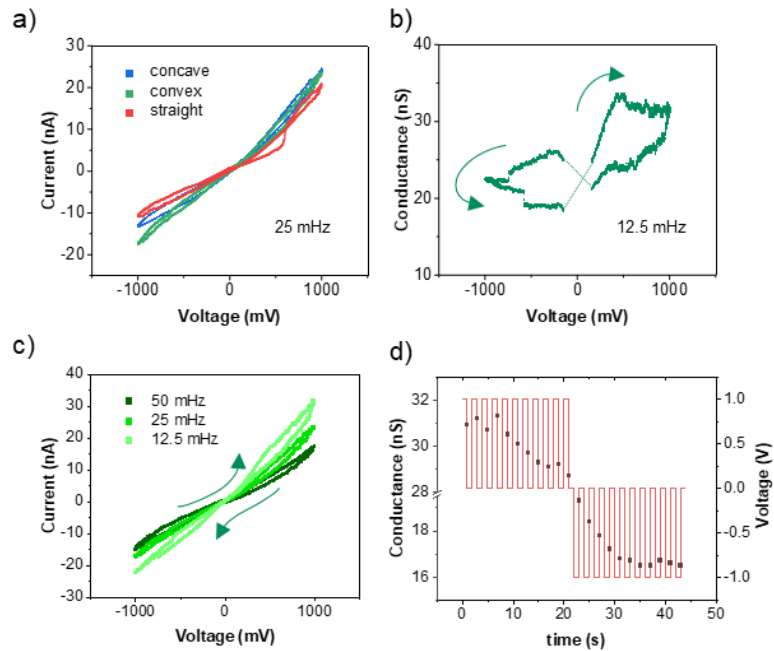

**Figure 5**. a) Memristive behavior of the SiO$_2$ 3D nanopores during cyclic ramp voltammetry in a 200/20 mM (tip side/base side) KCl gradient. b) The conductance of a conical pore of convex profile, showing the typical crossing of unipolar pores. The conductance curve is such that it decreases when the applied voltage increases, switching behavior when the sign of the voltage is changed. c) Memristive response depends on the cycling frequency. Here we show that the area of the hysteresis loop of a conical pore of convex profile increases with lower cycling frequencies. d) change in the conductivity (black) of the 50 nm concave pore as a series of 1s voltage pulses (red) is applied. These pores show a "learning" behavior, where the conductivity decreases as more pulses are applied.

**Conclusions**

In conclusion, we reported a fabrication route that allows to obtain nanopores with different conical shapes that consist fully of dielectric oxides. This is a key advancement in terms of nanopore shape and material, since it avoids typical shortcomings of polymer nanopores that are contamination of the analytes, low stability under harsh conditions and over long-time scales, and limited refractive index and dielectric constant that result in lower performance in optics and electronics. In this sense, our approach integrates the advantages of glass-pipette nanopores that are only available on the macro scale to on-chip technologies that enable parallel processing by micro arrays of nanopores, and their integration in microfluidics. The chemical robustness and high dielectric constant of the metal oxides that are available for ALD also open the way to integrate nanoplasmonics[69] and Mie-tronics with such on-chip nanopore platform. We demonstrated the excellent properties of the conical dielectric oxide nanopores in ionic current rectification, where they allowed to reach high rectification ratios even for relatively large pore diameters of 50-70 nm, and supporting our experiments with simulations elucidated the shape effects on the ionic current flow. We further demonstrated improved performance in electrical detection of DNA translocation of the conical nanopores compared to cylindrical ones, and connected the asymmetry in ionic current flow and conductance to the hysteresis that we observed in cyclic voltammetry. These insights showed that asymmetric shapes such as concave and convex nanopores are favorable for achieving hysteresis and memristive behavior, which allows to design nanopores for ionic memristor devices that have the best possible properties for the targeted applications, for example in neuromorphic computing. Therefore our work opens up new possibilities for exploring the physics and applications of conical nanopores, such as sensing[70], sequencing[71,72], nanofluidics[73], nanoelectronics[74], and energy harvesting[75]. With the addition of a metallic layer, the versatility of our method allows for the design of tunable plasmonic antennas for enhanced optical techniques[69,76]. Furthermore, exploiting the ability to fabricate 3D nanopores in different materials with high refractive index makes possible the design and integration of all dielectric nanoresonators, extending the applications of 3D nanopores into Mie-tronics[77,78].

**Methods and materials**

Ion beam and electron beam characterization were carried out by milling nanopores onto the $Si_3N_4$ membrane, coated with Microposit S1813 photoresist (spin-coated at 4000 rpm and baked at 95°C for 4 min) using $Ga^{+2}$ ions (Focused Ion Beam) with a currents ranging from 24 nA down to 7 pA, in Helios Nanolab 650 FEI. A series of concentric disks with diminishing diameters pattern was designed to fabricate the conical nanostructures. This same equipment was employed to inspect the morphology and dimensions of the nanopores using Scanning Electron Microscopy, and to perform EDX analyses and cross sections of the structures.

In Atomic Layer Deposition, a layer of the desired thickness (20 – 30 nm) of $SiO_2$, $Al_2O_3$, $HfO_2$, or $TiO_2$ was deposited using thermal and plasma processes at a constant temperature of 80°C. Subsequently, the sample was immersed in an Acetone bath for 5 minutes for development, and cleaned with Iso-Propyl Alcohol. For photoresist (PR) removal, the sample underwent UV light exposure for 1 minute to break the crosslinking in the hardened photoresist. This was followed by a low-power ICP-RIE oxygen plasma treatment for sample cleaning, utilizing 30W vertical power and 500W horizontal power. Finally, the sample was annealed at 650°C for one hour, employing a heating ramp of 3°C/minute and a cooling rate of 1°C/minute (15.4 hours in total).

*Ionic current measurements*

KCl was prepared at 10 mM concentration, at pH=7, using ultrapure MilliQ water, with resistivity of 18.2 MΩ·cm, as a solvent. All the solutions were prepared and used at room temperature and filtered with 200 nm-mesh Sartorius® filters.

In the rectification experiments KCl was prepared at 10 mM concentration at pH=7, using ultrapure MilliQ water, with resistivity of 18.2 MΩ·cm, as a solvent. The solutions were prepared and used at room temperature and filtered with 200 nm-mesh Sartorius® filters.

The electrical characterization was conducted using the e-NanoPore Reader (e-NPR) from Elements SRL with a selected frequency of 1.25 kHz for data acquisition. Ag/AgCl quasi-reference electrodes were prepared by immersing two Ag wires (0.5 mm diameter, 99.9% trace metals basis by Sigma Aldrich) in 5% sodium-hypochlorite (Sigma Aldrich)[58,79]. The experimental protocol for collecting Ionic Current Rectification (ICR) measures from all nanopores involved a staircase-shaped cyclic voltammetry. The voltage was changed by 100 mV every 20 seconds, ranging from 1 V to -1 V and then reversed. This procedure recorded the equivalents of reduction and oxidation branches, following the IUPAC convention[80]. The related electric current was the average of values registered in the last 1 second to avoid oscillations[81]. The measurements were conducted using the electrode facing the bottom (base) side of the nanopore as the anode (- electrode) and the reference electrode. The electrode facing the tip of the conical nanopore served as the cathode (+ electrode) and the working electrode.

*Nanopore wetting procedure*

Our[82] rectification experiment incorporates an efficient wetting procedure consisting of two key steps: plasma treatment and nanopore wetting. For the plasma treatment, the entire assembled fluidic cell, containing the nanopore chip, is exposed to 5 minutes of $O_2$ plasma at a power of 100 W. Subsequently, the fluidic cell and nanopore assembly are wetted using a solution comprising a 50:50%vol mixture of IPA and oxygen peroxide. Before each experiment, three injections of the wetting solution are administered, allowing the solution to be in contact with the microfluidic circuit for 5 minutes each time. Following this, two injections of the measured electrolyte are conducted – the first to rinse the wetting solution and the second for actual measurement. This wetting procedure has proven to yield reliable and effective Ionic Current Rectification (ICR) measurements.

*DNA translocations*

1μM λ-DNA translocations in 1 M KCl under applied voltages between V = +600 mV and -600 mV, employing a 20 kHz low-pass Bessel filter to discern event characteristics amidst background noise.

*Memristive behavior*

Cyclic voltammetry was conducted under different frequencies from 10 Hz to 1 mHz )where the frequency is the inverse of the duration of a single cycle) with a 1,2 to -1,2 V ramp, using 20 mM KCl on the tip side and 200 mM KCl on the base side of the conical nanopores. The same conditions were used for measuring the change in conductance as a series of pulses with 1s width and 1V amplitude was carried out, followed by a -1V pulse series[83].

ASSOCIATED CONTENT

**Supporting Information**.

brief description (file type, i.e., PDF)

AUTHOR INFORMATION

**Corresponding Author**

*denis.garoli@unimore.it; roman.krahne@iit.it

**Author Contributions**

GL conceived the fabrication method, performed the fabrication and prepared the manuscript, AS and DM performed the numerical simulations, AD, SW, GP and AG performed the memristor characterization, RK and DG supervised the work.

ACKNOWLEDGMENT

The authors thank the European Union under the Horizon 2020 Program, FET-Open: DNA-FAIRYLIGHTS, Grant Agreement 964995, the HORIZON-MSCA-DN-2022: DYNAMO, grant Agreement 101072818. The authors thank Clean Room Facility of IIT.

REFERENCES

1. Kasianowicz, J. J., Brandin, E., Branton, D. & Deamer, D. W. *Characterization of Individual Polynucleotide Molecules Using a Membrane Channel*. *Biophysics* vol. 93 https://www.pnas.org (1996).

2. Xue, L. *et al.* Solid-state nanopore sensors. *Nat Rev Mater* **5**, 931–951 (2020).

3. Hu, R., Tong, X. & Zhao, Q. Four Aspects about Solid-State Nanopores for Protein Sensing: Fabrication, Sensitivity, Selectivity, and Durability. *Advanced Healthcare Materials* vol. 9 Preprint at https://doi.org/10.1002/adhm.202000933 (2020).

4. Bocquet, L. Nanofluidics coming of age. *Nat Mater* **19**, 254–256 (2020).

5. Stein, D., Kruithof, M. & Dekker, C. Surface-charge-governed ion transport in nanofluidic channels. *Phys Rev Lett* **93**, (2004).

6. Turker Acar, E., Buchsbaum, S. F., Combs, C., Fornasiero, F. & Siwy, Z. S. *A P P L I E D S C I E N C E S A N D E N G I N E E R I N G Biomimetic Potassium-Selective Nanopores*. https://www.science.org (2019).

7. Di Muccio, G., Morozzo Della Rocca, B. & Chinappi, M. Geometrically Induced Selectivity and Unidirectional Electroosmosis in Uncharged Nanopores. *ACS Nano* **16**, 8716–8728 (2022).

8. Chernev, A. *et al.* Nature-Inspired Stalactite Nanopores for Biosensing and Energy Harvesting. *Advanced Materials* **35**, (2023).

9. Ramirez, P., Gómez, V., Cervera, J., Mafe, S. & Bisquert, J. Synaptical Tunability of Multipore Nanofluidic Memristors. *Journal of Physical Chemistry Letters* **14**, 10930–10934 (2023).

10. Designed Research; Y, X.-Y. K. T. & Performed Research; Y, P. L. T. Bioinspired nervous signal transmission system based on two-dimensional laminar nanofluidics: From electronics to ionics. *PNAS* **117**, (2005).

11. Dekker, C. Solid-state nanopores. *Nat Nanotechnol* **2**, 209–215 (2007).

12. Zhao, X., Qin, H., Tang, M., Zhang, X. & Qing, G. Nanopore: Emerging for detecting protein post-translational modifications. *TrAC Trends in Analytical Chemistry* **173**, 117658 (2024).

13. Bleidorn, C. Third generation sequencing: technology and its potential impact on evolutionary biodiversity research. *Syst Biodivers* **14**, 1–8 (2016).

14. Doricchi, A. *et al.* Emerging Approaches to DNA Data Storage: Challenges and Prospects. *ACS Nano* vol. 16 17552–17571 Preprint at https://doi.org/10.1021/acsnano.2c06748 (2022).

15. Thakur, M. *et al.* High durability and stability of 2D nanofluidic devices for long-term single-molecule sensing. *NPJ 2D Mater Appl* **7**, (2023).

16. He, Y., Tsutsui, M., Zhou, Y. & Miao, X. S. Solid-state nanopore systems: from materials to applications. *NPG Asia Materials* vol. 13 Preprint at https://doi.org/10.1038/s41427-021-00313-z (2021).

17. Liang, S. *et al.* Noise in nanopore sensors: Sources, models, reduction, and benchmarking. *Nanotechnology and Precision Engineering* **3**, 9–17 (2020).

18. Yi, W. *et al.* Solid-State Nanopore/Nanochannel Sensing of Single Entities. *Topics in Current Chemistry* vol. 381 Preprint at https://doi.org/10.1007/s41061-023-00425-w (2023).

19. Chen, Q. & Liu, Z. Fabrication and applications of solid-state nanopores. *Sensors (Switzerland)* vol. 19 Preprint at https://doi.org/10.3390/s19081886 (2019).

20. Healy, K., Schiedt, B. & Morrison, A. P. Solid-state nanopore technologies for nanopore-based DNA analysis. *Nanomedicine* **2**, 875–897 (2007).

21. Tseng, S., Lin, S. C., Lin, C. Y. & Hsu, J. P. Influences of cone angle and surface charge density on the ion current rectification behavior of a conical nanopore. *Journal of Physical Chemistry C* **120**, 25620–25627 (2016).

22. Chuang, P. Y. & Hsu, J. P. Influence of shape and charged conditions of nanopores on their ionic current rectification, electroosmotic flow, and selectivity. *Colloids Surf A Physicochem Eng Asp* **658**, (2023).

23. Cruz-Chu, E. R., Aksimentiev, A. & Schulten, K. Ionic cürrent rectification through silica nanopores. *Journal of Physical Chemistry C* **113**, 1850–1862 (2009).

24. Wen, C., Zeng, S., Li, S., Zhang, Z. & Zhang, S. L. On Rectification of Ionic Current in Nanopores. *Anal Chem* **91**, 14597–14604 (2019).

25. Trivedi, M. & Nirmalkar, N. Ion transport and current rectification in a charged conical nanopore filled with viscoelastic fluids. *Sci Rep* **12**, (2022).

26. Ai, Y., Zhang, M., Joo, S. W., Cheney, M. A. & Qian, S. Effects of electroosmotic flow on ionic current rectification in conical nanopores. *Journal of Physical Chemistry C* **114**, 3883–3890 (2010).

27. Wang, D. *et al.* Transmembrane potential across single conical nanopores and resulting memristive and memcapacitive ion transport. *J Am Chem Soc* **134**, 3651–3654 (2012).

28. He, Y., Tsutsui, M., Zhou, Y. & Miao, X. S. Solid-state nanopore systems: from materials to applications. *NPG Asia Materials* vol. 13 Preprint at https://doi.org/10.1038/s41427-021-00313-z (2021).

29. Paulo, G. *et al.* Hydrophobically gated memristive nanopores for neuromorphic applications. *Nat Commun* **14**, (2023).

30. Lee, J., Du, C., Sun, K., Kioupakis, E. & Lu, W. D. Tuning Ionic Transport in Memristive Devices by Graphene with Engineered Nanopores. *ACS Nano* **10**, 3571–3579 (2016).

31. Spende, A. *et al.* TiO2, SiO2, and Al2O3 coated nanopores and nanotubes produced by ALD in etched ion-track membranes for transport measurements. *Nanotechnology* **26**, (2015).

32. Fragasso, A., Schmid, S. & Dekker, C. Comparing Current Noise in Biological and Solid-State Nanopores. *ACS Nano* vol. 14 1338–1349 Preprint at https://doi.org/10.1021/acsnano.9b09353 (2020).

33. Chen, K. *et al.* Super-Resolution Detection of DNA Nanostructures Using a Nanopore. *Advanced Materials* **35**, (2023).

34. Roman, J. *et al.* Solid-State Nanopore Easy Chip Integration in a Cheap and Reusable Microfluidic Device for Ion Transport and Polymer Conformation Sensing. *ACS Sens* **3**, 2129–2137 (2018).

35. Gadaleta, A. *et al.* Sub-additive ionic transport across arrays of solid-state nanopores. *Physics of Fluids* **26**, (2014).

36. De Angelis, F. *et al.* 3D hollow nanostructures as building blocks for multifunctional plasmonics. *Nano Lett* **13**, 3553–3558 (2013).

37. Zhu, L. *et al.* Growth behavior of Ir metal formed by atomic layer deposition in the nanopores of anodic aluminum oxide. *Dalton Transactions* **51**, 9664–9672 (2022).

38. Chen, P. *et al.* Atomic layer deposition to fine-tune the surface properties and diameters of fabricated nanopores. *Nano Lett* **4**, 1333–1337 (2004).

39. Carlsen, A. T., Zahid, O. K., Ruzicka, J., Taylor, E. W. & Hall, A. R. Interpreting the conductance blockades of DNA translocations through solid-state nanopores. *ACS Nano* **8**, 4754–4760 (2014).

40. Garoli, D., Zilio, P., De Angelis, F. & Gorodetski, Y. Helicity locking of chiral light emitted from a plasmonic nanotaper. *Nanoscale* **9**, 6965–6969 (2017).

41. Garoli, D., Zilio, P., Gorodetski, Y., Tantussi, F. & De Angelis, F. Beaming of Helical Light from Plasmonic Vortices via Adiabatically Tapered Nanotip. *Nano Lett* **16**, 6636–6643 (2016).

42. Knoops, H. C. M. *et al.* Atomic Layer Deposition of Silicon Nitride from Bis(tert-butylamino)silane and N2 Plasma. *ACS Appl Mater Interfaces* **7**, 19857–19862 (2015).

43. Zhang, X. Y. *et al.* Deposition and characterization of rp-ald sio2 thin films with different oxygen plasma powers. *Nanomaterials* **11**, (2021).

44. Zelentsov, S. V., Zelentsova, N. V., Kolesov, A. N., Bogatyreva, L. A. & Mashtakov, I. A. Enhancing the dry-etch durability of photoresist masks: A review of the main approaches. *Russian Microelectronics* **36**, 40–48 (2007).

45. West, A., Van Der Schans, M., Xu, C., Cooke, M. & Wagenaars, E. Fast, downstream removal of photoresist using reactive oxygen species from the effluent of an atmospheric pressure plasma Jet. *Plasma Sources Sci Technol* **25**, (2016).

46. Le, Q. T. *et al.* Modification of Photoresist by UV for Post-Etch Wet Strip Applications. *Solid State Phenomena* **145–146**, 323–326 (2009).

47. Bauer, J., Crook, C. & Baldacchini, T. *3D PRINTING A Sinterless, Low-Temperature Route to 3D Print Nanoscale Optical-Grade Glass*. https://www.science.org.

48. Dingemans, G., van Helvoirt, C. A. A., Pierreux, D., Keuning, W. & Kessels, W. M. M. Plasma-Assisted ALD for the Conformal Deposition of SiO 2 : Process, Material and Electronic Properties . *J Electrochem Soc* **159**, H277–H285 (2012).

49. Aristanti, Y., Supriyatna, Y. I., Masduki, N. P. & Soepriyanto, S. Effect of calcination temperature on the characteristics of TiO2 synthesized from ilmenite and its applications for photocatalysis. in *IOP Conference Series: Materials Science and Engineering* vol. 478 (Institute of Physics Publishing, 2019).

50. Kruk, S. & Kivshar, Y. Functional Meta-Optics and Nanophotonics Govern by Mie Resonances. *ACS Photonics* vol. 4 2638–2649 Preprint at https://doi.org/10.1021/acsphotonics.7b01038 (2017).

51. Barsukova, M. G. *et al.* Magneto-Optical Response Enhanced by Mie Resonances in Nanoantennas. *ACS Photonics* **4**, 2390–2395 (2017).

52. Cervera, J., Schiedt, B., Neumann, R., Mafá, S. & Ramírez, P. Ionic conduction, rectification, and selectivity in single conical nanopores. *Journal of Chemical Physics* **124**, (2006).

53. Kiy, A. *et al.* Highly Rectifying Conical Nanopores in Amorphous SiO2Membranes for Nanofluidic Osmotic Power Generation and Electroosmotic Pumps. *ACS Appl Nano Mater* **6**, 8564–8573 (2023).

54. Gubbiotti, A. *et al.* Electroosmosis in nanopores: computational methods and technological applications. *Adv Phys X* **7**, 2036638 (2022).

55. Bruus, H. *Theoretical Microfluidics*. vol. 18 (Oxford Academic, 1997).

56. Kubeil, C. & Bund, A. The role of nanopore geometry for the rectification of ionic currents. *Journal of Physical Chemistry C* **115**, 7866–7873 (2011).

57. Apel, P. Y., Blonskaya, I. V., Orelovitch, O. L., Ramirez, P. & Sartowska, B. A. Effect of nanopore geometry on ion current rectification. *Nanotechnology* **22**, (2011).

58. Niedzwiecki, D. J., Chou, Y. C., Xia, Z., Thei, F. & Drndić, M. Detection of single analyte and environmental samples with silicon nitride nanopores: Antarctic dirt particulates and DNA in artificial seawater. *Review of Scientific Instruments* **91**, (2020).

59. Lanzavecchia, G. *et al.* Plasmonic Photochemistry as a Tool to Prepare Metallic Nanopores with Controlled Diameter for Optimized Detection of Single Entities. *Adv Opt Mater* (2023) doi:10.1002/adom.202300786.

60. Hou, Y. & Hou, X. Bioinspired nanofluidic iontronics. *Science* vol. 373 628–629 Preprint at https://doi.org/10.1126/science.abj0437 (2021).

61. Liu, W. *et al. Bioinspired Carbon Nanotube-Based Nanofluidic Ionic Transistor with Ultrahigh Switching Capabilities for Logic Circuits*. *Sci. Adv* vol. 10 https://www.science.org (2024).

62. Noh, Y. & Smolyanitsky, A. Memristive Response and Capacitive Spiking in Aqueous Ion Transport through Two-Dimensional Nanopore Arrays. *Journal of Physical Chemistry Letters* **15**, 665–670 (2024).

63. Roy, K., Jaiswal, A. & Panda, P. Towards spike-based machine intelligence with neuromorphic computing. *Nature* **575**, 607–617 (2019).

64. Marković, D., Mizrahi, A., Querlioz, D. & Grollier, J. Physics for neuromorphic computing. *Nature Reviews Physics* vol. 2 499–510 Preprint at https://doi.org/10.1038/s42254-020-0208-2 (2020).

65. Chua, L. If it's pinched it's a memristor. *Semicond Sci Technol* **29**, 104001 (2014).

66. Ramirez, P., Portillo, S., Cervera, J., Bisquert, J. & Mafe, S. Memristive arrangements of nanofluidic pores. *Phys Rev E* **109**, 44803 (2024).

67. Kamsma, T. M. *et al.* Brain-inspired computing with fluidic iontronic nanochannels. *Proceedings of the National Academy of Sciences* **121**, e2320242121 (2024).

68. Kamsma, T. M., Boon, W. Q., ter Rele, T., Spitoni, C. & van Roij, R. Iontronic Neuromorphic Signaling with Conical Microfluidic Memristors. *Phys Rev Lett* **130**, 268401 (2023).

69. Li, W. *et al.* Enhanced Optical Spectroscopy for Multiplexed DNA and Protein-Sequencing with Plasmonic Nanopores: Challenges and Prospects. *Analytical Chemistry* vol. 94 503–514 Preprint at https://doi.org/10.1021/acs.analchem.1c04459 (2022).

70. Storm, A. J., Chen, J. H., Ling, X. S., Zandbergen, H. W. & Dekker, C. Fabrication of solid-state nanopores with single-nanometre precision. *Nat Mater* **2**, 537–540 (2003).

71. Lee, K. *et al.* Recent Progress in Solid-State Nanopores. *Advanced Materials* **30**, 1704680 (2018).

72. Dorey, A. & Howorka, S. Nanopore DNA sequencing technologies and their applications towards single-molecule proteomics. *Nature Chemistry* vol. 16 314–334 Preprint at https://doi.org/10.1038/s41557-023-01322-x (2024).

73. Pérez-Mitta, G., Toimil-Molares, M. E., Trautmann, C., Marmisollé, W. A. & Azzaroni, O. Molecular Design of Solid-State Nanopores: Fundamental Concepts and Applications. *Advanced Materials* **31**, 1901483 (2019).

74. Krapf, D. *et al.* Fabrication and Characterization of Nanopore-Based Electrodes with Radii down to 2 nm. *Nano Lett* **6**, 105–109 (2006).

75. Feng, J. *et al.* Single-layer MoS2 nanopores as nanopower generators. *Nature* **536**, 197–200 (2016).

76. Garoli, D., Yamazaki, H., MacCaferri, N. & Wanunu, M. Plasmonic Nanopores for Single-Molecule Detection and Manipulation: Toward Sequencing Applications. *Nano Letters* vol. 19 7553–7562 Preprint at https://doi.org/10.1021/acs.nanolett.9b02759 (2019).

77. Koshelev, K. *et al. Subwavelength Dielectric Resonators for Nonlinear Nanophotonics*. https://www.science.org.

78. Kivshar, Y. The Rise of Mie-tronics. *Nano Letters* vol. 22 3513–3515 Preprint at https://doi.org/10.1021/acs.nanolett.2c00548 (2022).

79. Xia, Z. *et al.* Silicon Nitride Nanopores Formed by Simple Chemical Etching: DNA Translocations and TEM Imaging. *ACS Nano* **16**, 18648–18657 (2022).

80. Elgrishi, N. *et al.* A Practical Beginner's Guide to Cyclic Voltammetry. *J Chem Educ* **95**, 197–206 (2018).

81. Yang, C., Hinkle, P., Menestrina, J., Vlassiouk, I. V. & Siwy, Z. S. Polarization of Gold in Nanopores Leads to Ion Current Rectification. *Journal of Physical Chemistry Letters* **7**, 4152–4158 (2016).

82. Giacomello, A. & Roth, R. Bubble formation in nanopores: a matter of hydrophobicity, geometry, and size. *Advances in Physics: X* vol. 5 Preprint at https://doi.org/10.1080/23746149.2020.1817780 (2020).

83. Robin, P. *et al. Long-Term Memory and Synapse-like Dynamics in Two-Dimensional Nanofluidic Channels*. https://www.science.org (2023).